\documentclass[nofootinbib,twocolumn,showpacs,preprintnumbers,amsmath,amssymb, showkeys, noshowpacs]{revtex4}
\usepackage{fancyhdr}
\usepackage{indentfirst}
\usepackage{hhline}
\usepackage{wrapfig, float}
\usepackage{array,longtable,tabularx}
\usepackage{graphicx}% Include figure files
\usepackage{dcolumn}% Align table columns on decimal point
\usepackage{bm}% bold math
\usepackage{subfigure}
\usepackage[small]{caption2}
\usepackage{csquotes}
\usepackage{footnote,enumitem}
\usepackage[referable]{threeparttablex}
\usepackage{boxhandler}
\usepackage{color}
\usepackage{hyperref}
\usepackage{textcomp}% qu
\renewcommand{\captionlabeldelim}{.}
\hypersetup{linkcolor=blue, citecolor=blue, urlcolor=blue, colorlinks=true}%, pdfborderstyle={/S 1}}
\begin{document}

\title{The analysis of heat capacity of MnGe metallic helimagnet}

\author{M.A.~Anisimov$^{1, 2}$}\email{anisimov.m.a@gmail.com} \author{A.V.~Bokov$^{2}$}\author{A.V.~Semeno$^{1, 2}$}
\author{V.A.~Sidorov$^{1}$}\author{A.V.~Tsvyashchenko$^{2}$}

\affiliation{$\phantom{x}^1$- Prokhorov General Physics Institute of the Russian Academy of Sciences, 38 Vavilov Street, 119991, Moscow, Russia}

\affiliation{$\phantom{x}^2$- Vereshchagin Institute for High Pressure Physics of RAS, 14 Kaluzhskoe Shosse, 142190 Troitsk, Russia}

\begin{abstract}
Zero-field heat capacity of metallic helimagnet MnGe was analyzed based on the results of resistivity decomposition published previously by our group for the same crystal. Current procedure allowed identifying along with ($i$) electronic ($\tilde{\gamma}$ $\approx$ 7~mJ/mol$\cdot$K$^2$) and ($ii$) phononic ($\Theta$$_\textmd{D}$ $\approx$ 350~K) components ($iii$) the additional term, caused by the presence of spin fluctuations (SFs). The last contribution was found to exist in a wide range of temperatures in both paramagnetic (PM) and magnetically ordered states. However, its amplitude appears to be significantly lower in comparison with phononic component. The obtained value of spin fluctuation temperature $\theta$$_\textmd{{sf}}$(MnGe) $\approx$ 330~K correlates well with previous estimations, as well as with results of various experiments, which predict the existence of SFs in MnGe at least up to 250 $-$ 300~K.
\end{abstract}

\keywords{heat capacity analysis, Debye contribution, spin fluctuation temperature, MnGe, helimagnets}
\maketitle

%\textbf{1.}
\section*{1. Introduction}\label{Sec.1}
Helical magnets of B20 family with noncentrosymmetric crystal structure (sp. gr. $P2_13$) have attracted considerable attention in condensed-matter physics. Here long wavelength helical modulations are stabilized due to the competition between several different magnetic interactions, including Dzyaloshinskii-Moriya and ferromagnetic (FM) ones \cite{1, 2}. Anisotropic exchange, dipolar interactions, and cubic crystal field anisotropy also play role in determining the characteristics of the helical structure.

Among other members of B20 series, monogermanide MnGe, which is isostructural analogue of MnSi, stands apart, because of its unusual properties. It exhibits a helical spin structure with a propagation vector \textbf{\textit{k}} = (0, 0, $\zeta$) in reciprocal lattice units with $\zeta$ = 0.107 just below Neel point $T_\textmd{N}$ = 170~K \cite{3, 4}. The parameter $\zeta$ increases upon cooling and then locks-in to the value 0.167 at $T_{\textmd{com}}$ = 32~K. Then it does not change below $T_{\textmd{com}}$ and helical period becomes commensurate with the lattice \cite{3, 5}. At the same time above $T_{\textmd{com}}$ a phase separation (PS) was proposed in a wide range of temperatures ($T_{\textmd{com}}$ $\leq$ $T$ $\leq$ $T_\textmd{N}$) \cite{6, 7}. Possible three-dimensional skyrmion lattice (SKL) or $A$-phase was inferred from small-angle neutron scattering (SANS) \cite{8} and from Lorenz transmission electron microscopy experiments \cite{9}. This circumstance distinguishes MnGe from MnSi, for which the two-dimensional triangular SKL was discussed in literature. Moreover, alternating skyrmions and antiskyrmions have been proposed for MnGe in zero-field \cite{10}. However, recent neutron studies questioned SKL conception \cite{11} at least for the host MnGe, so magnetic ground state of this system is the subject of debates up to now.

The value of helix wave vector for MnGe is $k_\textmd{s}$ = 2.2~nm$^{-1}$ (for comparison for MnSi $k_\textmd{s}$ = 0.35~nm$^{-1}$), the magnetic moment per Mn atom at 2~K is equal to 1.9~$\mu_\textmd{B}$, which is 5 times larger than that one for MnSi (0.4~$\mu_\textmd{B}$). MnGe is also characterized by the shortest helix pitch (29~$\textmd{\AA}$ versus 180~$\textmd{\AA}$ for MnSi) and by high enough value of critical field $H_\textmd{c}$ = 15~T, needed to transform helix structure into FM state. Besides, a transition from a high-spin to a low-spin state was observed for MnGe under applied pressure around $P$ = 6~GPa \cite{4}. Recent theoretical calculations have shown that the Dzyaloshinskii-Moriya constant is unexpectedly equal to zero for MnGe \cite{12}. All these arguments allowed the authors \cite{13} proposing the scenario with a relatively small value of Dzyaloshinskii-Moriya interaction (DMI) for this object and therefore more important role of effective RKKY-interaction for stabilization of helical structure. There are also theoretical approaches predicting the formation of a helical structure without taking into account DMI, but considering instead only the symmetry of the system.

At the same time skyrmion issues do not limit the broad spectrum of physical phenomena in MnGe, which includes also a large topological Hall \cite{14}, and topological Nernst effects \cite{14, 15}. Moreover, the study of spin fluctuations (SFs) is of great fundamental interest as well. Here SFs play significant role in the ground state formation, including the stabilization of intermediate PS state. Recent decomposition of zero-field resistivity showed that SFs survives with cooling down to low temperatures \cite{16}.

Here we performed the analysis of zero-field heat capacity of this extraordinary object. These data were measured previously by our group with co-authors and published in \cite{17}. Moreover, the $C$($T$) dependence of Pauli-paramagnetic counterpart CoGe \cite{17}, which is also classified in literature as Weyl semimetal \cite{18}, will be used here as auxiliary data. Although specific heat of these systems have been previously presented in a series of works \cite{17, 19, 20}, no detailed analysis has been suggested.

%_______________________Fig 1 ________________________________
\begin{figure}[!t]
\begin{center}
\includegraphics[width = 8cm]{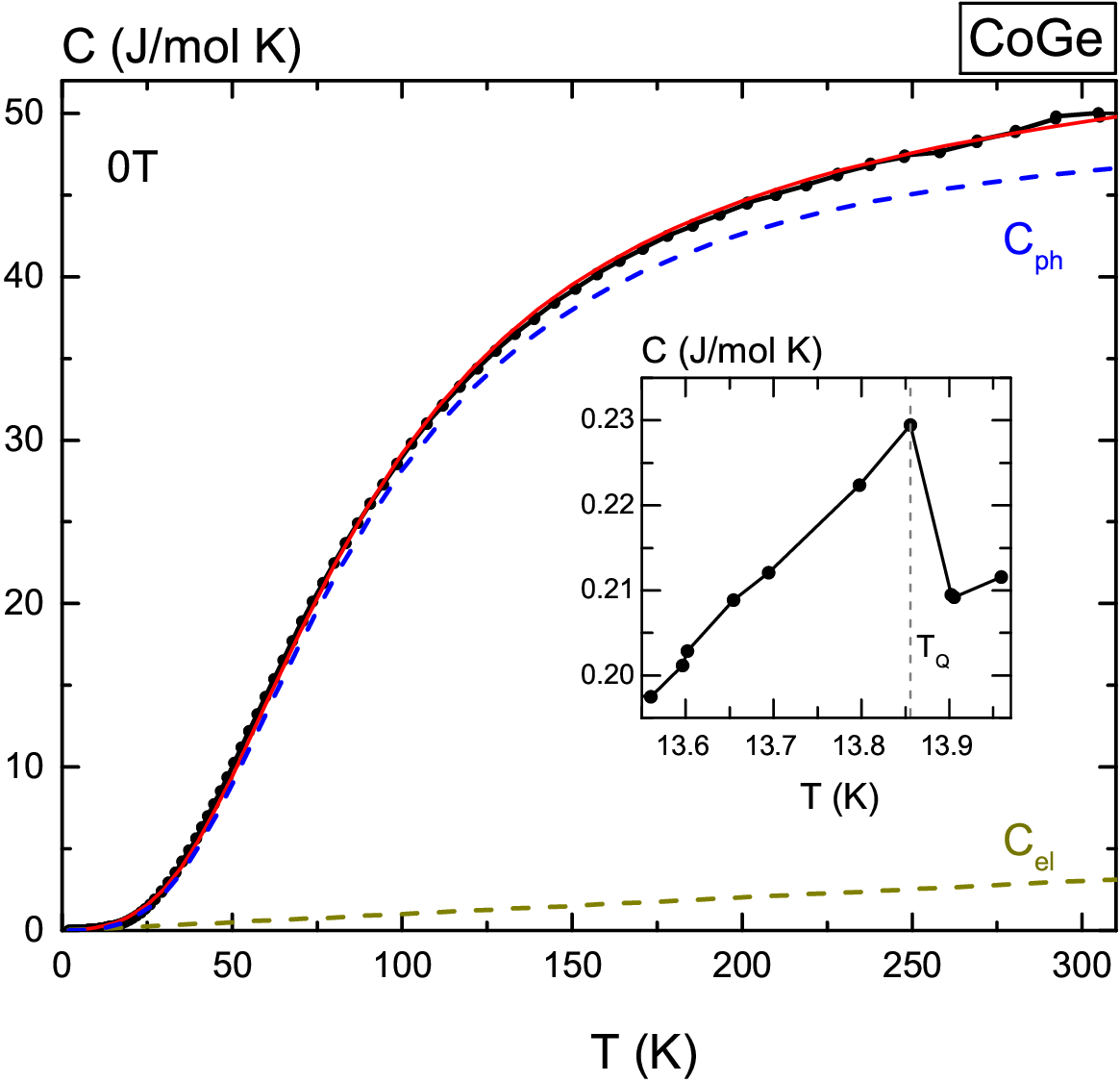}
  \caption{(Color online). Temperature evolution of zero-field heat capacity $C$($T$) of CoGe. Solid line displays the summary approximation by Eq.(\hyperref[Eq1]{1}) or model I. Dashed lines of different colors present the decomposition analysis with the estimation of electronic ($C_{\textmd{el}}$) and phononic ($C_{\textmd{ph}}$) components.  Parameters of fitting are given in text. The inset illustrates zoom of low-temperature anomaly at $T_\textmd{Q}$ = 13.85~K.}\label{FigX1}

   \end{center}
\end{figure}
%_______________________Fig 1 ________________________________

%%%%%%%%%%%%%%%%%%%%%%%%%%%%%%%%%%%%%%%%%%

\section*{2. Results and discussion}\label{Sec2}
\begin{figure*}[!htpb]
\includegraphics[width = 16.5cm]{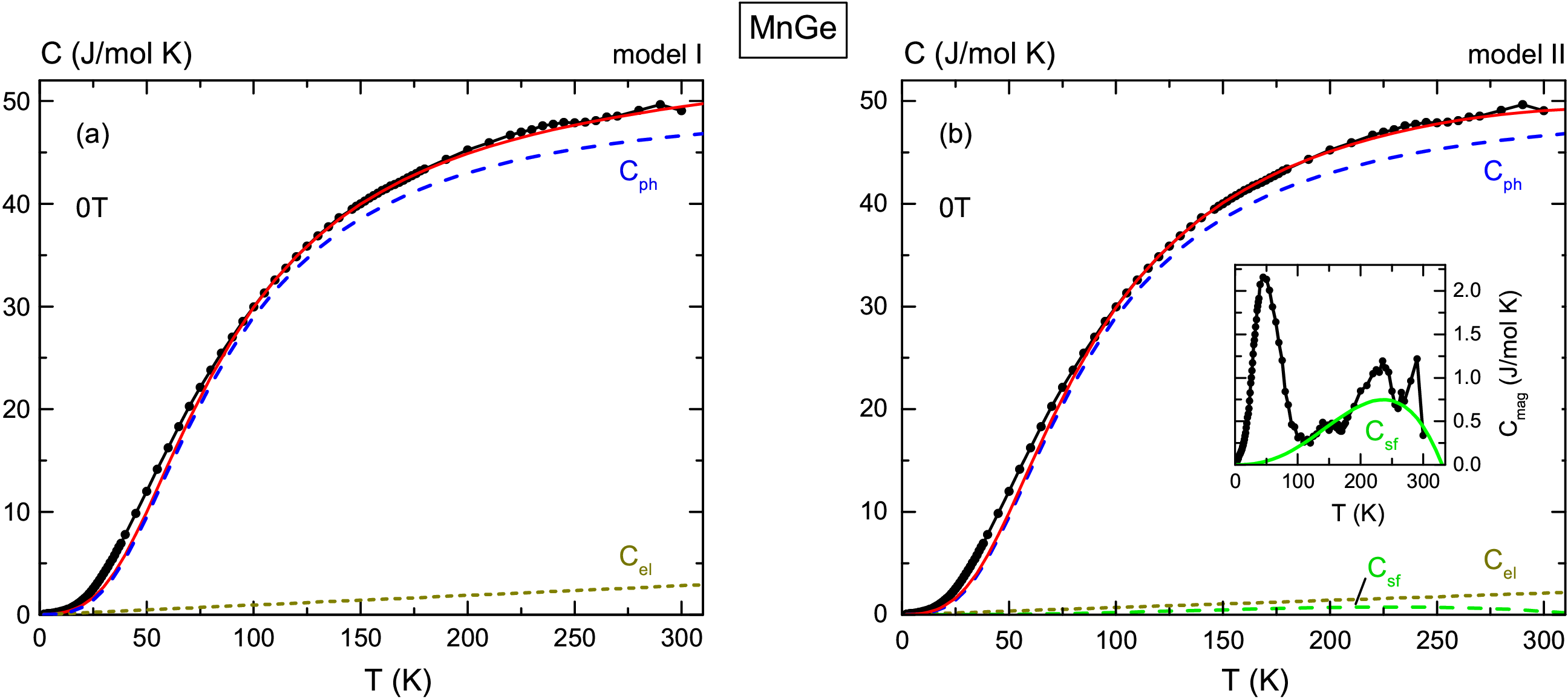}
   \parbox{16cm}{\caption{(Color online). Temperature evolution of zero-field heat capacity $C$($T$) of MnGe. Solid line displays the summary approximation (a) by Eq.(\hyperref[Eq1]{1}) or model I and (b) with additional contribution $C_{\textmd{sf}}$($T$), caused by spin fluctuations (or model II). Dashed lines present the decomposition analysis with estimation of electronic, phononic and [in panel (b) only] additional spin-fluctuation components. Parameters of fitting are given in text. The inset in panel (b) illustrates magnetic term $C_\textmd{{mag}}$($T$) and the fit by Eq.(\hyperref[Eq3]{3}), see Sect.\textbf{2.3}.}}\label{FigX2}
\end{figure*}

\textbf{2.1.}\label{Sec2p1}  Figures \hyperref[FigX1]{1}-\hyperref[FigX2]{2} present temperature evolution of zero-field heat capacity $C$($T$) of CoGe and MnGe, respectively. Despite the fact that several anomalies were detected on heat capacity curves of monosilicide analogue MnSi, including a narrow peak at $T_\textmd{c}$ = 29~K and broad shoulder preceding it \cite{21}, no visible peculiarities were registered on obtained data of MnGe in the vicinity of Neel point ($T_\textmd{N}^{\rho(T)}$ = 160 $-$ 170~K, \cite{16}) and the signal decreases monotonically upon cooling (Fig.\hyperref[FigX2]{2}). In the case of CoGe there is an additional weak low-$T$ feature at $T_\textmd{Q}$ = 13.85~K ($\Delta$$C$ = 19~mJ/mol~K). The fragment of the curve in enlarged scale is depicted for this material in the inset of Fig.\hyperref[FigX1]{1}. Such feature was studied in details in \cite{20}, including the measurements with smaller temperature step. According to \cite{20}, mentioned anomaly survives with applied magnetic field at least up to 9~T. It was interpreted as a fingerprint of quadrupole-order-driven commensurate-incommensurate phase transition in CoGe. We will not focus on it below.

As it was shown in \cite{22}, the implementation of standard procedure, when the ratio $C$/$T$ = $f$($T^2$) is fitted by linear law, $i.e.$ $\gamma_0$ + $\beta$$T^2$ without a proper estimation of the additional dominating magnetic component, may lead to significant errors, especially for magnetic systems, such as MnGe. As a rule, routine algorithm gives overestimated value of Sommerfeld coefficient $\gamma_0$ = $\pi^2/3g(E_\textmd{F})k_\textmd{B}T$ [where $g(E_\textmd{F})$ denotes the density of states at Fermi level], which does not correspond to the asymptotic of original curve and may result to a negative sign of heat capacity in decomposition as an artifact. Of course, in general case Sommerfeld coefficient can depend on temperature and, accordingly, enhance with cooling \cite{23}-\cite{28}\footnotemark{}. \footnotetext{Here we used the notation $\gamma_0$ = $C_{\textmd{el}}$/$T$$|_{T \longrightarrow 0}$ for low-$T$ part of electronic coefficient, while $\tilde{\gamma}$ corresponds to high-$T$ plateau.}Sometimes it may reach drastic amplitude ($\gamma_0$ $>$ 400~mJ/mol$\cdot$K$^2$) at low temperatures as it is realized in heavy fermion (HF) compounds \cite{24}. (Maximal values of $\gamma_0$ $\sim$ 10$^4$~mJ/mol$\cdot$K$^2$ are detected for so-called superheavy fermions objects \cite{25}-\cite{28}). But such effect always requires correct consideration of the magnetic part, which has not been done to date. In particular, the overestimated values of Sommerfeld coefficient of $\gamma_{\textmd{inc}}$(GdB$_6$) $\approx$ 632~mJ/mol$\cdot$K$^2$, $\gamma_{\textmd{inc}}$(PrB$_6$) $\approx$ 280~mJ/mol$\cdot$K$^2$, $\gamma_{\textmd{inc}}$(HoB$_6$) $\approx$ 207~mJ/mol$\cdot$K$^2$  were earlier reported for antiferromagnetic metals GdB$_6$, PrB$_6$ and HoB$_6$ in literature \cite{29}-\cite{31}. According to arbitrary criterion \cite{24, 32}, such materials should be categorized as classical ($\gamma$ $>$ 400~mJ/mol$\cdot$K$^2$) or moderate ($\gamma$ = 100 $-$ 400~mJ/mol$\cdot$K$^2$) HF systems. At the same time, other characteristics of these objects show no signs of HF behavior. For example, zero-field resistivity versus temperature demonstrates metallic character \cite{33}-\cite{35} with the absence of Kondo-type anomalies. Quantum oscillations and optical experiments restrict effective mass of electrons for these materials by the limit of $m^*$(GdB$_6$) = 0.6~$m_0$ \cite{36} and $m^*$(PrB$_6$) = 0.28 $-$ 2.52~$m_0$ \cite{33, 37}. The discussion in \cite{22} shows that ferromagnetic semimetal EuB$_6$ with Sommerfeld coefficient $\gamma_{\textmd{inc}}$(EuB$_6$) $\approx$ 820~mJ/mol~K$^2$ obtained using routine procedure also should be considered as classical HF compound. But as it was pointed in \cite{22} this object has a low concentration of electrons $\sim$ 10$^{19}$~cm$^{-3}$ for EuB$_6$, corresponding to effective concentration as $n$/$n_{4f}$ = 10$^{-3}$ \cite{38}. For this reason electronic component was neglected for EuB$_6$ ($\gamma_0$ = $\tilde{\gamma}$ = 0) in $C$($T$) decomposition \cite{22}.

%_______________________Fig 2 ________________________________
%_______________________Fig 2 ________________________________

%%%%%%%%%%%%%%%%%%%%%%%%%%%%%%%%%%%%%%%%%%

Similar situation is realized for Debye temperature ($\Theta_{\textmd{D}}$), which is usually estimated in standard algorithm from coefficient $\beta$ as $\Theta_{\textmd{D}}$ = $\left(12\pi^4nR/5\beta\right)^{1/3}$. For example, the comparison of the values of $\Theta_{\textmd{D}}$ parameter obtained for nonmagnetic reference system LaB$_6$ within routine procedure and from the simulation of entire curve of both heat capacity $C$($T$) and resistivity $\rho$($T$) gives considerable discrepancy as 364~K for the first case \cite{31} and 1160~K for the last one \cite{23, 39, 40}. Due to the need of renormalization in $R$B$_6$ set this difference becomes even greater for nonmagnetic topological insulator candidate YbB$_6$ as 274~K \cite{31} versus 1160~K, correspondingly \cite{22}. The procedure of renormalization often does not work and the correctness of applying of renormalization factor $k$ = [$\Theta_{\textmd{D}}$(LaB$_6$)/$\Theta_{\textmd{D}}$($R$B$_6$)]$^3$ = [$M$($R$B$_6$)/$M$(LaB$_6$)]$^{3/2}$ (where $M$ $-$ is a molar mass)  \cite{41} requires the implementation of $T^3$ expansion for the phonon part of heat capacity (Debye law), which is realized only at low temperatures \cite{42}. Therefore the correct evaluation of both $\gamma$ and $\Theta_{\textmd{D}}$ parameters is very important and the error in either can significantly shift the values of other variables used in the analysis.

\textbf{2.2.}\label{Sec2p2} For this reason in initial iteration it is convenient to approximate high temperature region of experimental curve by the sum of electronic ($C_{\textmd{el}}$) and phononic ($C_{\textmd{ph}}$) components (model I)

%%%%%%%%%%%%%%%%%%%%%%%%%%%%%%%%%%%%%%%
\begin{equation}\label{Eq1}
C = \tilde{\gamma} T + C_{ph}.
\end{equation}

\noindent
The second term in Eq.(\hyperref[Eq1]{1}) is defined by the relation

\begin{equation}\label{Eq2}
C_{\textmd{ph}} = 9nR\left(\frac{T}{\Theta_\textmd{D}}\right)^3 \int_{0}^{\Theta_\textmd{D}/T}e^x x^4 (e^{x} - 1)^{-2}dx,
\end{equation}

\noindent
where $R$ is the universal gas constant, and $n$ $-$ structural coefficient. The last parameter describes the number of atoms in a formula unit, $i.e.$ $n$ = 2 for CoGe and MnGe. Such a method allows controlling the high temperature slope of simulated curve by varying of the value of $\tilde{\gamma}$,  since $C_{ph}$ term with the Debye temperature close to 300~K exhibits the saturation behavior with heating. As result, parameters of approximation were estimated as $\tilde{\gamma}$ $\approx$ 10.1~mJ/mol$\cdot$K$^2$, $\Theta_{\textmd{D}}$ $\approx$ 360~K (31~meV) and $\tilde{\gamma}$ $\approx$ 9.2~mJ/mol$\cdot$K$^2$ and $\Theta_{\textmd{D}}$ $\approx$ 350~K (30.2~meV) for CoGe and MnGe materials, respectively (see solid lines in Figs.\hyperref[FigX1]{1}, \hyperref[FigX2]{2a}). The obtained values of $\Theta_{\textmd{D}}$ correlate well with each other and with results of zero-field resistivity decomposition, performed for Mn$_{1-x}$Fe$_x$Ge family [$\Theta_{\textmd{D}}^{\rho(T)}$ = 300~K (25.85~meV)], as well as for MnSi helimagnet [$\Theta_{\textmd{D}}^{\rho(T)}$ = 350~K] in \cite{16}. Despite the fact that the conception of Weyl semimetal is used for CoGe \cite{18}, the nonzero Sommerfeld coefficient was obtained here (dark yellow dashed line in Fig.\hyperref[FigX1]{1}).

Various methods of heat capacity analysis are discussed in literature, including those, in which Debye temperature depends on $T$ \cite{43}-\cite{45}. In this case all features of phonon spectrum are reflected in temperature dependence of the function $\Theta_{\textmd{D}}$ = $f$($T$). There are also more exotic techniques, in which the contribution of each of sublattices is calculated independently by separate Debye function \cite{46}-\cite{48}. However, in current work we choose the variant with single parameter $\Theta_{\textmd{D}}$ to be fixed as a constant. In such a method Mn/Co and Ge sublattices are modeled identically and estimated value of $\Theta_{\textmd{D}}$ represents the average contribution of each of sublattices.

%Reply
The existence of quasilocal mode in MnGe/CoGe should be discussed separately, although its presence is much less obvious for B20 systems in comparison with the conception of two Debye sublattices. Unlike $R$B$_6$ (RE hexaboride) and $R$B$_{12}$ (RE dodecaboride) classes, where rigid covalent framework of boron complexes produces Debye term (DT) [continuous spectrum in phonon density of states (PDOS)] and quasilocal vibrations of RE ions are responsible for Einstein contribution ($\delta$-function in PDOS) \cite{39} so that both components may be considered as independent, in B20 case everything is not so evident. In particular, it is unclear which elements/lattices should contribute to quasilocal mode. Of course, the system with complex phonon spectrum such as LaB$_6$/LuB$_{12}$ or MnGe may have many anomalies. However, its primitive description inevitably requires simplification. Sometimes there may be different variants of solution. For example, except currently accepted for $R$B$_6$ and $R$B$_{12}$ classes model \cite{39} some alternative algorithms use the decomposition of both resistivity and heat capacity, taking into account only the sum of several Einstein components \cite{Tessier, Lorz}. Such procedure makes it possible to obtain high precision approximation and to estimate PDOS as well \cite{Tessier, Lorz}. At the same time it is not clear what the minimal number of the modes with different Einstein temperatures ($\Theta_{\textmd{E}}$) should be considered (see the detailed discussion in \cite{22, Anisimovarxiv}). Moreover, some of low frequencies $\omega$$_{\textmd{Ei}}$ may be erroneously interpreted as Einstein modes, while they may appear due to the presence of boron vacancies (defect mode) \cite{22, 23}. In any case the exclusion of DT should be justified. Here one may point on the results of Ref. \cite{Bolotina}, where the decomposition of zero-field resistivity of LuB$_{12}$, including the separation of normal and Umklapp processes was carried out using only single Einstein mode. Proof of inaccuracy of such procedure is the artificial depression of the value of residual resistivity, the estimate of which in the analysis turns out significantly lower than the experimental one, as it was mentioned in \cite{Anisimovarxiv}.

For multiparameter approximation even the choice of suitable algorithm does not guarantee the obtaining of correct parameters of phonon spectrum. In particular, sometimes the fit gives reversed values of $\Theta_{\textmd{D}}$ and $\Theta_{\textmd{E}}$ for some members of $R$B$_6$ family \cite{EuB6arxiv, Tanaka}, so that the condition $\Theta_{\textmd{E}}$ $>$ $\Theta_{\textmd{D}}$ is realized. The last fact contradicts to the results of \cite{39}. For this reason the self-consistent analysis of heat capacity together with corresponding approximation of resistivity and/or the mean square displacement of atoms is typically applied to compare the results and exclude possible artifacts \cite{39}. Here we used previous analysis of zero-field resistivity of MnGe (\cite{16}, see also Sect.\textbf{2.3}), which shows no signs of characteristic markers of quasi-local mode \cite{39, Cooper}, similar to $C$($T$) data presented in Figs.\hyperref[FigX1]{1}-\hyperref[FigX2]{2}. Besides, the fit of Pauli-paramagnetic counterpart CoGe by Eq.(\hyperref[Eq1]{1}) describes well  experimental curve $C$($T$) without Einstein component (solid line in Fig.\hyperref[FigX1]{1}). Interestingly, for isostructural MnSi the relative length change $\Delta$$L/L$ = $f$($T$) was approximated by single DT ($\Theta_{\textmd{D}}$ = 350~K) in the wide range of temperatures of paramagnetic phase \cite{16}. Therefore for MnGe system with high enough Neel temperature the quasi-local mode was not considered, in accordance with Occam's razor principle. Another aspect to discuss is that general conditions for deriving Debye model do not assume the absence of the center of inversion in the crystal. It may be proposed that for this case the condition for dispersion relation $\omega$($k = 0$) = 0 is violated. However, more accurate adaptation of Debye model to B20 systems requires separate detailed theoretical consideration and will be discussed elsewhere.

%end Reply

\textbf{2.3.}\label{Sec2p3}  As a rule the magnetic contribution to heat capacity is estimated in literature by subtracting the experimental curve of nonmagnetic counterpart. Sometimes using this method may lead to the appearance of artifacts due to strong renormalization of the phonon spectrum. For example, the reduction of Einstein temperature was reported for RE hexaborides and dodecaboride families \cite{63}-\cite{67}. In particular, it starts from $\Theta_\textmd{E}$ $\approx$ 140 $-$ 152~K \cite{23, 39} for referenced system LaB$_6$ and decreases down to $\Theta_\textmd{E}$(YbB$_6$) $\approx$ 92~K and $\Theta_\textmd{E}$(GdB$_6$) $\approx$ 91~K for both divalent and trivalent members, respectively \cite{23, 63}. In the set HoB$_{12}$ $-$ LuB$_{12}$ (with nonmagnetic analogue LuB$_{12}$) the spread of values is slightly smaller $\Theta_\textmd{E}$(HoB$_{12}$) $\approx$ 206~K versus $\Theta_\textmd{E}$(LuB$_{12}$) $\approx$ 160~K, as it follows from extended X-ray absorption fine structure (EXAFS) spectroscopy experiments \cite{67}.

In contrast to CoGe, for which model I completely describes experimental curve (Fig.\hyperref[FigX1]{1}), for MnGe there is an additional magnetic contribution $C_{\textmd{mag}}$($T$). The last one was obtained after the subtraction of the sum of electronic and phononic components from experimental curve [$C_{\textmd{mag}}$ = $C_{\textmd{exp}}$ $-$ $C_{\textmd{ph}}$ $-$ $C_{\textmd{el}}$], see symbols in the inset in Fig.\hyperref[FigX2]{2b}. It is clear that the function $C_{\textmd{mag}}$($T$) exhibits two features. One of them is the peak centered at $T_{\textmd{max}}$ $\approx$ 45~K, which turns out to be lower than the temperature of lock-in to commensurate structure $T_{\textmd{com}}^{\rho(T)}$ = 59 $-$ 63~K, estimated for Mn$_{1-x}$Fe$_x$Ge family ($x$ = 0, 0.02) from zero-field resistivity derivative \cite{16}. Further heating above $T_{\textmd{max}}$ leads to the formation of a broad anomaly at $T_{\textmd{max2}}$ = 235~K, which amplitude is considerably depressed in comparison with the first peculiarity. Significant difference between these values allows us proposing different nature of them. We suggest that the last one (the feature at $T_{\textmd{max2}}$) appears due to SFs contribution. Indeed, the analysis of zero-field resistivity $\rho$($T$) shows that the component $\rho_{\textmd{sf}}$($T$), caused by the scattering of electrons on localized spin fluctuations \cite{49, 50}, dominates over phononic term in both paramagnetic and magnetically ordered states of MnGe \cite{16}. It is known from literature \cite{51}-\cite{56}, that SFs contribute to heat capacity as

\begin{equation}\label{Eq3}
C_{\textmd{sf}} = A_{\textmd{sf}}\left(\frac{T}{\theta_{\textmd{sf}}}\right)^3ln\left(T/\theta_{\textmd{sf}} \right),
\end{equation}

\noindent
where $A_{\textmd{sf}}$ is the amplitude constant and $\theta_{\textmd{sf}}$ is characteristic temperature, related to spin fluctuations. In this case $C_{\textmd{sf}}$ component should be added to the sum in Eq.(\hyperref[Eq1]{1}), (model II). As result a better convergence of the fit with experimental data is observed at high-$T$ range. The values of mentioned above parameters were computed as $A_{\textmd{sf}}$ $\approx$ $-$6.1~J/mol$\cdot$K and $\theta_{\textmd{sf}}$ $\approx$ 330~K (27.6~meV), correspondingly. The appearance of a new term with close to $\Theta_{\textmd{D}}$ value of spin-fluctuation temperature shifts only Sommerfeld coefficient as $\tilde{\gamma}$ $\approx$ 7~mJ/mol$\cdot$K$^2$, see solid line in Fig.\hyperref[FigX2]{2b}. Here we used $C_{\textmd{sf}}$ only as a small correction to the model I fit [Eq.(\hyperref[Eq1]{1})], since in the analysis of heat capacity of MnGe the phonon contribution $C_{\textmd{ph}}$ turns out to be the dominating one (blue dashed line in Figs.\hyperref[FigX2]{2a}-\hyperref[FigX2]{2b}). As result, $C_{\textmd{sf}}$ affects only on the electronic component. Note also that the extruded value of $\theta_{\textmd{sf}}$ $\approx$ 330~K correlates well with result $\theta_{\textmd{sf}}^{\rho(T)}$ $\approx$ 300~K obtained in \cite{16}. The specific of the estimation of $C_{\textmd{sf}}$ term is that its asymptotic is depressed down to zero at $\theta_{\textmd{sf}}$ as it follows from logarithmic law in Eq.(\hyperref[Eq3]{3}), see solid  line in the inset of Fig.\hyperref[FigX2]{2b}.

Spin fluctuation component to heat capacity was predicted theoretically in \cite{51}-\cite{53}. Moreover, $C_{\textmd{sf}}$ was also discussed for many systems, including metallic antiferromagnet GdCu$_6$ \cite{54}, Laves phase NdRh$_2$ \cite{56}, spin-fluctuation systems UAl$_2$, UCo$_2$, UPt$_3$, U$_{1-x}$Th$_x$Al$_2$ \cite{55}, \cite{57}-\cite{59}; amorphous iron-zirconium ($a$-Fe$_x$Zr$_{100-x}$) alloys \cite{60}; copper-rich Cu-Ni alloys \cite{61}, and even for topological insulator candidate SmB$_6$ \cite{62}. However, for the last object the applicability of such approach is under debates, since SFs were not reported for SmB$_6$ in literature. As a rule, the using of Eq.(\hyperref[Eq3]{3}) to heat capacity analysis requires also the decomposition of magnetic susceptibility or resistivity to prove the approach. In particular, SFs contribute as $T^2$ at low temperatures for the first characteristic. On the contrary, River-Zlatic model  \cite{49, 50} predicts the evolution of resistivity with temperature from quadratic $\rho_{sf}$ $\sim$ $T^2$ and linear $\sim$ $T$ trajectories at low and intermediate temperatures at $T$ $<$ $\theta_{\textmd{sf}}$ to logarithmic law $\sim$ $ln$($T$) above $\theta_{\textmd{sf}}$. Exactly the similar behavior of $\rho$($T$) curve was observed for MnGe in \cite{16}. Interestingly, SFs not always depressed with external magnetic field as it was mentioned in \cite{16, 57}.

Note also that ferromagnetic and chiral spin fluctuations were experimentally detected in paramagnetic and in magnetically ordered states of MnGe. For this compound PS is realized in intermediate phase, when slow and rapid chiral SFs associated with long-range and short range ordered helices coexist within a large temperature interval below Neel point (100~K $<$ $T$ $<$ $T_\textmd{N}$ \cite{6}). Besides, rapid spin fluctuations survive in PM state of MnGe at temperatures at least up to 250 $-$ 300~K \cite{7}, that correlate well with the estimated values of $\theta_{\textmd{sf}}$. Yet, the relative fluctuation fraction of Mn moments, deduced from the analysis of M$\ddot{o}$ssbauer spectra, shows that the phase separation appears only above $\sim$ 100~K \cite{6}. This boundary may be shifted down to 50~K, according to $\mu$SR studies \cite{7} or even to 32~K in experiments of Modulation of Intensity Emerging with Zero Effort spectroscopy \cite{5}.

\begin{figure}[!t]
\begin{center}
\includegraphics[width = 8cm]{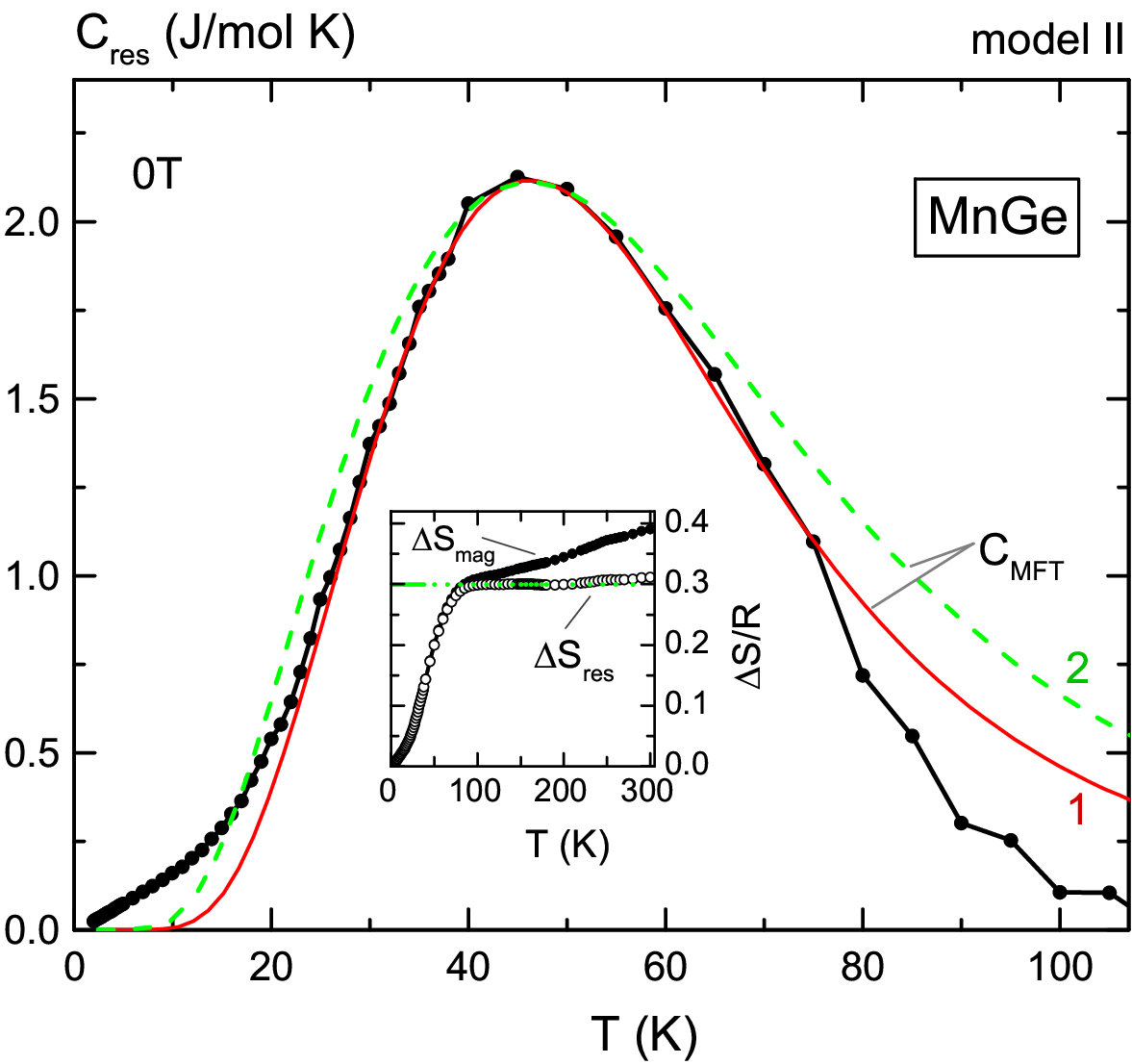}
  \caption{(Color online).  Residual magnetic contribution $C_{\textmd{res}}$ versus temperature, estimated for MnGe within model II. Lines (1)-(2) display approximation computed in the framework of mean-field theory, Eqs.(\hyperref[Eq4]{4})-(\hyperref[Eq7]{7}). Among them the first one (1) corresponds to the case of $S$ = 1/2 ($\mathfrak{R}$ = 1, $H_0$ = 0.97~a.u.), while the second one (2) shows the fit with $S$ = 1 ($\mathfrak{R}$ = 1, $H_0$ = 1.59~a.u.). The inset illustrates temperature evolution of magnetic entropy change $\Delta$$S_{\textmd{mag}}$($T$) (solid symbols) together with the $\Delta$$S_{\textmd{res}}$($T$) function (open symbols) computed from Eq.(\hyperref[Eq8]{8}). Here dashed line corresponds to the value 0.43$R$$ln$(2), see Sect.\textbf{2.4} for details.
  }\label{FigX3}

   \end{center}
\end{figure}

\textbf{2.4.}\label{Sec2p4}  Now we will focus on the low-temperature anomaly in $C_{\textmd{mag}}$($T$) dependence. For ease of analysis we subtracted SFs term [Eq.(\hyperref[Eq3]{3})] from magnetic component, $i.e.$ $C_{\textmd{res}}$ = $C_{\textmd{mag}}$ $-$ $C_{\textmd{sf}}$, see Fig.\hyperref[FigX3]{3}. As result the function $C_{\textmd{res}}$($T$) goes down to zero at $T_\textmd{z}$ = 105~K, as it is depicted in Fig.\hyperref[FigX3]{3}. Note also that our data do not exhibit low-$T$ upturn, caused by nuclei contribution.

%_______________________Fig 3 ________________________________
%_______________________Fig 3 ________________________________

%%%%%%%%%%%%%%%%%%%%%%%%%%%%%%%%%%%%%%%%%%

In general case the analysis of magnetic component to heat capacity requires computer simulations, which take into account the presence of DMI in the system, as it was made early for MnSi in \cite{68, 69}. However, MnGe is characterized by a low value of Dzyaloshinskii-Moriya constant \cite{12}. For this reason we performed the approximation of zero-field $C_{\textmd{res}}$($T$) in the framework of mean-field theory (MFT) \cite{70}

\begin{equation}\label{Eq4}
	C_\textmd{{MFT}} = -\frac{3R}{2}\frac{S}{(S+1)}\frac{d(m^2)}{dt},
\end{equation}

% {d\left(\frac{T}{T_\textmd{C}}\right)}\left(\frac{d~m^2}{dt}\right)^2
\noindent
where $t$ = $T$/$T_{\textmd{C}}$ and $m$ = $M$/$M_0$ correspond to reduced temperature and magnetization, respectively. In Eq.(\hyperref[Eq4]{4}) $M_0$ denotes the saturation magnetization and $S$ $-$ spin number. To determine the magnetization $M$, a numerical solution of the system is required

\begin{equation}\label{Eq5}
	m = B_\textmd{S}(x) = \frac{2S+1}{2S} coth\left(\frac{2S+1}{2S} x\right) - \frac{1}{2S} coth\left(\frac{x}{2S}\right),
\end{equation}
\noindent\begin{equation}\label{Eq6}
	B = \mu_0 H_0 + \left(\lambda - N_\textmd{d}\right) \mu_0 M,%\nonumber,
\end{equation}
\noindent
\begin{equation}\label{Eq7}
	x = \frac{g \mu_\textmd{B} S}{k_\textmd{B} T} \left(\mu_0 H_0 + \left(\lambda - N_\textmd{d}\right) \mu_0 M\right), %\nonumber,
\end{equation}

\noindent
where  $x$ =($g$$\mu_{\textmd{B}}$$S$)/($k_{\textmd{B}}$$T$)$H$ is the dimensionless argument in Brillouin function $B_{\textmd{S}}$($x$), $H_0$ $-$ applied magnetic field, $\lambda$ $-$ mean-field constant and $N_\textmd{d}$ demagnetization factor (0 $\leq$ $N_\textmd{d}$ $\leq$ 1). In the analysis we used the parameter $\mathfrak{R}$ = $\lambda$ $-$ $N_\textmd{d}$ for convenience. Two lines in Fig.\hyperref[FigX3]{3} illustrate the approximation of $C_{\textmd{res}}$ by MFT using different values of spin number $S$, including (1) $S$ = 1/2 and (2) $S$ = 1. The application of the MFT necessitates the presence of a low magnetic field $H_0$ even for the case of zero-field data. Increasing the value of $H_0$ leads to smooth of the fitting curve \cite{23, 56, 71}. As result, other parameters of analysis were changed as (1) $\mathfrak{R}$ = 1, $H_0$ = 0.97~a.u. and (2) $\mathfrak{R}$ = 1, $H_0$ = 1.59~a.u., respectively. According to Fig.\hyperref[FigX3]{3}, fit 1 meets $C_{\textmd{res}}$($T$) data quite well at least near the maximum. In contrast, curve (2) is significantly broadened. In both cases a sufficiently large value of $H_0$ is needed. Of course, current analysis oversimplifies real processes occurring in MnGe system. For more detailed study of magnetic contribution further computer simulations are required. However, this will need the choice of an adequate three dimensional model.

Zero-field magnetic entropy change ($\Delta$$S_{\textmd{mag}}$) was calculated in the framework of the expression

\begin{equation}\label{Eq8}
\Delta S_\textmd{{mag}}(T) = \int C_\textmd{{mag}}(T)/T dT.
\end{equation}
\noindent
where  $C_{\textmd{mag}}$($T$) corresponds to summary magnetic contribution (see Sect.\textbf{2.3}). The function $\Delta$$S_{\textmd{res}}$ = $f$($T$) was also computed (open symbols in the inset of Fig.\hyperref[FigX3]{3}). In this case the $C_{\textmd{res}}$($T$) term was used in the numerator of Eq.(\hyperref[Eq8]{8}) instead of $C_{\textmd{mag}}$. In order to avoid the violation of the third law of thermodynamics, we extrapolated $C_{\textmd{mag}}$($T$) and $C_{\textmd{res}}$($T$) data down to 0~K using a parabolic polynomial $\sim$ $aT$ $+$ $bT^2$ with zero shift along the vertical axis. Similar technique was applied previously to the analysis of zero-field magnetic entropy change in ferromagnetic EuB$_6$ in \cite{23, 72} and Laves phase NdRh$_2$ \cite{56}. The application of such procedure assumes that the nuclei contribution $C_{\textmd{nuc}}$($T$) to heat capacity may be neglected. Indeed, according to previous studies $C_{\textmd{nuc}}$($T$) is responsible for the rise of heat capacity as $\sim$ $T^{-2}$ at low temperatures (the tail of Schottky asymptotic). At the same time, nuclei component usually gives a small correction to the summary magnetic entropy change and may be ignored.

Our estimates indicate that both functions $\Delta$$S_{\textmd{mag}}$($T$) and $\Delta$$S_{\textmd{res}}$($T$) do not cross the line $R$$ln$(2$S$ $+$ 1) with $S$ = 1/2 [$i.e.$ $R$$ln$(2)] at least up to 300~K (solid and open symbols in the inset of Fig.\hyperref[FigX3]{3}, respectively). On the contrary, they demonstrate the tendency to saturation behavior in the vicinity of $T_\textmd{z}$ with maximal value as 0.45$R$$ln$(2) [0.43$R$$ln$(2)], respectively. The heating above $T_\textmd{z}$ leads to the the monotonic rise of $\Delta$$S_{\textmd{mag}}$($T$), while for $\Delta$$S_{\textmd{res}}$($T$) dependence the saturation survives. To our knowledge, similar effect with reduced amplitude of $\Delta$$S_{\textmd{mag}}$($T$) is realized also for host MnSi \cite{73}, for Mn$_{1-x}$Fe$_x$Si, Mn$_{1-x}$Co$_x$Si \cite{74} and Fe$_{0.8}$Co$_{0.2}$Si \cite{75} compositions as well as for recently synthesized helimagnet Fe$_{0.5}$Rh$_{0.5}$Si \cite{76}, ect. In literature it is common to attribute such rather small entropy release at the magnetic phase transition to weak itinerant-electron magnetism behavior \cite{74, 75}. This is also may point on strong magnetic fluctuations in the system. Indeed, SFs were detected in intermediate PS state and also in a wide range of paramagnetic phase of MnGe, according to various investigations \cite{6, 7}. As for MnSi, it was established in \cite{2, 77, 78} that this object undergoes two crossovers with three different regimes of SFs in PM vicinity of $T_\textmd{c}$. Among them ferromagnetic fluctuations are observed at high-$T$ up to 50~K \cite{77} that may explain the behavior of $\Delta$$S_{\textmd{mag}}$($T$) function for MnSi.

\section*{3. Conclusion}\label{Sec3}
To summarize, the original primitive  procedure of zero-field heat capacity decomposition was proposed for MnGe and CoGe materials. Our results allowed excluding  zero value of Sommerfeld coefficient for Pauli-paramagnetic counterpart CoGe ($\tilde{\gamma}$ $\approx$ 10.1~mJ/mol$\cdot$K$^2$), despite the concept of Weyl semimetal proposed for this compound in literature. Besides, the analysis performed for CoGe shows no need to consider the contribution of the quasi-local mode. Two models were suggested for MnGe. It was shown that spin-fluctuations contribution with $\theta_{\textmd{sf}}$(MnGe) $\approx$ 330~K together with electronic term ($\tilde{\gamma}$ $\approx$  7~mJ/mol$\cdot$K$^2$) and dominating phonon component [$\Theta_{\textmd{D}}$(MnGe) $\approx$ 350~K] leads to better fit of experimental data. Magnetic component  $C_{\textmd{res}}$($T$) was analyzed for MnGe in the framework of mean-field theory, which describes peak only with minimal spin number $S$ = 1/2.

Moreover, current work examines the aspects of heat capacity analysis and demonstrates in details that classical approach is inapplicable to the systems with magnetic order. Considering these conclusions one can question the validity of the results obtained for MnSi in literature. Indeed, according to our knowledge, the heat capacity of this object has not been measured in the temperature range 100 $-$ 300~K, which is the most important for correct decomposition as it was mentioned in Sect.\textbf{2.2}.

However, there are a number of limitations for the presented variant of approximation. For example, the estimation of SFs contribution for MnGe in the form of Eq.(\hyperref[Eq3]{3}) does not take into account the existence of chiral spin fluctuations. It is applied in literature for FM spin fluctuations \cite{54,59}, which were also detected for MnGe. In addition, magnetic and phononic components are considered here for MnGe as independent ones. Besides, the using of MFT to describe magnetic contribution oversimplifies real processes occurring in this extraordinary system. The next step requires computer simulations.

\section*{CRediT authorship contribution statement}
\textbf{M.A.~Anisimov}: Conceptualization, Formal analysis, Methodology, Visualization, Writing $-$ original draft, Writing $-$ review $\&$ editing. \textbf{A.V.~Bokov}: Resources. \textbf{A.V.~Semeno}: Resources. \textbf{V.A.~Sidorov}: Resources, Investigation. \textbf{A.V.~Tsvyashchenko}: Resources, Project administration, Funding acquisition, Writing $-$ review $\&$ editing.

\section*{ORCID iDs}
\begin{longtable}[t]{lcc}%\firsthline
M.A.~Anisimov & \quad \small{\href{https://orcid.org/0000-0002-8449-9894}{https://orcid.org/0000-0002-8449-9894}} \\
A.V.~Bokov & \quad\small{\href{https://orcid.org/0000-0001-8487-2719}{https://orcid.org/0000-0001-8487-2719}} \\
A.V.~Semeno & \quad\small{\href{https://orcid.org/0000-0002-6150-6086}{https://orcid.org/0000-0002-6150-6086}} \\
V.A.~Sidorov & \quad\small{\href{https://orcid.org/0000-0002-3861-0707}{https://orcid.org/0000-0002-3861-0707}} \\
A.V.~Tsvyashchenko & \quad\small{\href{https://orcid.org/0000-0002-8282-5107}{https://orcid.org/0000-0002-8282-5107}} \\
\end{longtable}

\section*{Declaration of competing interest}
The authors declare that they have no known competing financial interests or personal relationships that could have appeared to influence the work reported in this paper.

\section*{Data availability}
Data will be made available on request.

\section*{Acknowledgements}
The authors would like to thank V.N.~Krasnorussky, Dr. V.V.~Glushkov and Prof. S.V.~Demishev for useful discussions.

\section*{Funding}
This work was supported by a grant from the Russian Science Foundation No. 25-12-68013 (22-12-00008-$\pi$).

%%%% ----------------------------------------------------------------------

\end{document}